\documentstyle[aps,prl,floats,epsf]{revtex}  
\begin{document}                
\twocolumn[\hsize\textwidth\columnwidth\hsize\csname
 @twocolumnfalse\endcsname
%
%
%
\title{Development of Magnetism in  
Strongly Correlated Cerium Systems: Non-Kondo Mechanism for
Moment Collapse}
\author{Eric M. Collins, Nicholas Kioussis, and Say Peng Lim}
\address{Department of Physics and Astronomy, California State University 
Northridge,Northridge, CA 91330-8268}
\author{Bernard R. Cooper}
\address{Department of Physics, West Virginia University, Morgantown, 
WV 26506-6315}
\maketitle
\begin{abstract}                
We present an {\it ab initio} based method which gives clear insight into
the interplay between the hybridization, the coulomb exchange, and 
the crystal-field interactions, as the degree of 4{\it f} localization 
is varied across a series of strongly correlated cerium systems.
The results for the ordered magnetic moments, magnetic structure, and ordering
temperatures are in excellent agreement with experiment, including
the occurence of a {\it moment collapse} of non-Kondo origin. 
In contrast, standard {\it ab initio} density functional calculations
fail to predict, even qualitatively, the trend of the unusual magentic
properties.

\end{abstract}

\vspace{+12pt}

]
%
\newpage
The difficulties and interest in treating strongly correlated 
electron systems, and the consequences of correlation 
effects on magnetic behavior in the transitional 4{\it f} or 
5{\it f} localization regime, 
provide one of the central problems of condensed matter physics.$^{1-3}$
The transitional regime behavior is neither atomiclike nor itinerant.
This gives rise to an extremely interesting range of phenomena, 
but also causes very great difficulties in treating the
theory of these phenomena adequately, especially in a way providing the
ability to predict the behavior of specific materials.$^{1-3}$
An adequate treatment requires treating the interelectronic coulomb 
interaction, i.e. the correlation effects, as constrained by 
exchange symmetry.$^{4-6}$  In this letter, we demonstrate an approach
for treating these difficulties in predicting the interesting and
complex behavior of an important series of cerium compounds.

The isostructural (rock-salt structure) series of the cerium
monopnictides CeX (X = P, As, Sb, Bi) and monochalcogenides (X=S, Se, Te)
have become prototype model systems for study,
because of their unusual magnetic properties.$^{7-13}$
This series of strongly correlated electron systems offers the opportunity
to vary systematically, through chemical pressure, the lattice constant
and the cerium-cerium separation on going down the pnictogen
or chalcogen column, and hence tailor the degree of 4{\it f} localization
from the strongly correlated limit in the heavier systems to the weakly
correlated limit in the lighter systems.$^{7-13}$
The calculated single-impurity Kondo temperature, T$_K$,
 presented below, is much smaller than the 
magnetic ordering temperature in these systems, 
 and hence this series lies in 
 the {\it magnetic} regime of the Kondo phase diagram.$^{14}$ 
Nevertheless, in this work we demonstrate that 
the sensitivity of the hybridization, 
coulomb exchange, and crystal-field
interactions with the chemical environment gives rise to a variety of
unusual and interesting magnetic properties across the series, 
in agreement with 
experiment, {\it including the occurence 
of a non-Kondo magnetic moment collapse}.


This class of cerium systems exhibits large magnetic anisotropy which
changes from the $<001>$ direction in the pnictides to the
$<111>$ direction in the chalcogenides. The low-temperature
ordered magnetic moment increases with increasing lattice constant
for the pnictides from 0.80$\mu_B$ in CeP to 2.1$\mu_B$ in 
CeSb and CeBi,$^{7-8}$
while it decreases with increasing lattice constant for the chalcogenides
from 0.57$\mu_B$ in CeS to 0.3$\mu_B$ in CeTe.$^{7-9}$
The {\it magnetic moment collapse} from CeSb to CeTe, with
both systems having about the
same lattice constant, is indicative of the sensitivity of the magnetic
 interactions to chemical environment.
The experimentally observed low-temperature structure 
in CeBi and CeSb
is the  
$<001>$ antiferromagnetic 
type IA ($\uparrow \uparrow \downarrow \downarrow)$, 
whereas in CeAs and CeP the structure is the  
$<001>$ antiferromagnetic type I ($\uparrow \downarrow$).$^{7,15}$  
The ordering temperature increases from 8K in CeP
to 26K in CeBi for the pnictides, whereas it decreases from
8.4K in CeS to an unusually low 2.2K in CeTe.$^{7-11}$
Another unusual feature of this series of cerium compounds 
is the large {\it suppression} of the crystal field (CF) splitting
of the Ce$^{3+}$ free-ion 4$f_{5/2}$ multiplet from
values expected from the behavior of the heavier
isostructural rare-earth pnictides or chalcogenides.$^{16}$
This can be understood$^{17}$ as arising from band-{\it f} hybridization
effects.
In both the cerium monopnictides and monochalcogenides,
 the CF splitting between the ${\Gamma_{7}}$ doublet and the
${\Gamma_{8}}$ quartet
decreases with increasing anion size,
 from 150 K for CeP to 10 K in CeBi and from 130 K for CeS to 30 K for CeTe,
and it is about the same for the same row in both series,
 a rather surprising result in view of the
additional valence electron on the chalcogen ion.$^{18}$
  Neutron scattering experiments have shown$^{19}$ that
the ${\Gamma_{7}}$-doublet
is the CF ground state in all the cerium pnictides and
chalcogenides.


In this paper we present material-predictive results from 
two {\it ab initio} based 
methods to study the change of magnetic properties  
 across this series of cerium systems.
The first, {\it ab initio} 
based, method  
gives clear insight into the role of the three pertinent interactions:
1) The band-{\it f} hybridization-induced inter-cerium magnetic 
coupling; 2) the corresponding effects of band-{\it f} coulomb exchange; and
3) the crystal-field interaction. This approach 
allows us also to understand the interplay between 
these interactions
as the degree of 4{\it f} localization is varied across the series.
The predictive calculations give results 
for the magnetic moments, magnetic structure, and ordering 
temperatures in excellent agreement with experiment. 
Thus, this approach allows to understand and predict
 a number of key features 
of observed behavior. 
 First, is the very low moment and low ordering 
temperature of the antiferromagnetism 
observed in CeTe,  
 an incipient heavy Fermion system. (For a review of theory 
and experimental behavior of heavy Fermion systems 
see references 2,3,20,21.)  
This {\it ab initio}-based method, 
described below, predicts the {\it magnetic moment and 
ordering temperature collapse}
from CeSb to CeTe, both systems having about the same lattice
constant but CeTe having an additional {\it p} electron.
The origin of the moment collapse is of non-Kondo origin.
The earlier work of Sheng and Cooper$^5$ showed that this 
magnetic ordering reduction is accurately predicted without
including any crystal-field effects. 
An erroneous statement appears in the recent review 
article by Santini {\it et al.}$^{22}$ stating that 
crystal-field effects played an important role in the 
calculated results of Sheng and Cooper.$^5$ This is incorrect, 
since crystal-field effects were {\it not} included in these calculations.
We show in this 
paper that including the crystal-field effects modifies
this behavior only quantitatively. 
Second, our results demonstrate that, while the band-{\it f} coulomb exchange
mediated interatomic 4{\it f}-4{\it f} interactions dominate 
the magnetic behavior for the heavier systems, 
which are more localized because of the larger Ce-Ce separation, 
the opposite is true for the lighter, more delocalized systems, 
where the hybridization-mediated coupling dominates the
magnetic behavior. This reflects the great sensitivity
of the relative importance of hybridization and coulomb 
exchange effects on magnetic ordering depending on the
degree of 4{\it f} localization.
Third, we show that 
for the lighter more delocalized systems the crystal-field
interactions are much larger than the inter-cerium interactions
and hence dominate the magnetic behavior.
Finally,  
 we predict the experimentally observed 
change of the ground-state magnetic 
structure from the  
$<001>$ antiferromagnetic
type IA ($\uparrow \uparrow \downarrow \downarrow)$ in CeBi and CeSb
 to the $<001>$ antiferromagnetic type I ($\uparrow \downarrow$) 
 in CeAs and CeP. 
On the other hand, the second {\it ab initio} method, based on 
density functional theory within the local density
approximation (LDA),$^{23,24}$ fails to predict, even qualitatively, 
the trend of magnetic properties in this series of strongly
correlated electron systems.

The first, {\it ab initio} based, method employs the 
degenerate Anderson lattice model which incorporates 
explicitly the hybridization and the coulomb exchange
 interactions on an equal footing$^{4,5}$  

\small
\begin{eqnarray*}
H &=&\sum_{k} \epsilon_{k} c_{k}^{+} c_{k}
    + \sum_{Rm} \epsilon_{m} f_{m}^{+}(R) f_{m}(R) \\
  & & + \frac{U}{2} \sum_{R,m \ne m'} n_{m}(R) n_{m'}(R) \\ 
  & & +\sum_{kmR} [ V_{km} e^{-i{\bf k \cdot R}} c_{k}^{+} f_{m}(R)
   + H.C. ] \\ 
  & &  - \sum_{kk'} \sum_{mm'R} J_{mm'}({\bf k,k'})
   e^{-i{\bf (k-k') \cdot R}} c_{k}^{+} f_{m}^{+}(R) c_{k'}
   f_{m'}(R).
\end{eqnarray*}
\vspace{-30pt}
\begin{equation}
\end{equation}
\normalsize
The parameters entering the model Hamiltonian, i.e., 
the band energies $\epsilon_k$, the {\it f}-state energy $\epsilon_m$, 
the on-site coulomb
repulsion U, the hybridization matrix elements, V$_{km}$, and the
band-{\it f} coulomb exchange
$J_{m m'}({\bf k, k'}) =
          \left< \phi_{k}^{*}({\it r}_{1}) \psi_{m}^{*}({\it r}_{2})
               \left|\frac{1}{{\it r}_{12}}\right|
          \psi_{m'}({\it r}_{1})\phi_{k'}({\it r}_{2})\right>$
are  evaluated on a wholly {\it ab initio} basis 
from non-spin polarized full potential linear muffin tin orbital$^{23}$ 
(FPLMTO) calculations.
Here, ${\it r}_{12}$ stands for $|{\it r}_{1}-{\it r}_{2}|$;
$\phi_{k}$ are the non-{\it f} basis states of the FPLMTO, and
$\psi_{m}$ are the localized {\it f} states.
Because the size of both the hybridization and coulomb exchange
matrix elements are much smaller ($\sim$ 0.1 eV) than the intraatomic
 coulomb interaction U (6eV), one can apply perturbation
theory and evaluate the anisotropic two-ion 6X6 
interaction  matrices$^{4,5}$, 
$E_{m_{1}m'_{1}}^{m_{2}m'_{2}}{\bf (R_{2}-R_{1})}$, 
which couple the 
 two {\it f}-ions. 
The exchange interactions have three contributions:
   the wholly band-{\it f} coulomb exchange
mediated interaction 
proportional to $J_{mm'}^{2}({\bf k,k'})$, the wholly hybridization-mediated
exchange interaction proportional to $V_{km}^{4}$,
and the cross term proportional to $V_{km}^{2} J_{mm'}({\bf k,k'})$.  
With the two-ion interactions having been determined, the low-temperature
magnetic moment and the ordering temperature can be determined by use
of a mean field calculation.$^{4,5,8}$
We have previously
 applied this {\it ab initio} based method to 
investigate the effect of hybridization-induced
cerium-cerium interactions$^{4,17}$ and the combined effect 
of both the hybridization
and coulomb induced interactions$^{5}$ on the magnetic properties of
the heavier cerium pnictides and chalcogenides (CeBi, CeSb, and CeTe).
However, these calculations did not take into account
the crystal field interaction and employed a warped muffin-tin LMTO
calculation for the parameters entering the model.
The excellent agreement found$^{5}$ with experiment for the low-temperature
magnetic moment and ordering
temperature is 
relatively unaffected by the CF interaction,  
because the CF interaction in the heavier
cerium systems is smaller than the two-ion exchange interactions.

The second method employs {\it ab initio} 
spin polarized electronic structure calculations
based on the FPLMTO method$^{23}$ 
using 1) only spin polarization,
with the orbital polarization included
only through the spin-orbit coupling, and 2) both  the spin and orbital
polarization polarization.$^{24}$
 In these calculations the 4{\it f} states are treated as band states.
The orbital polarization is
 taken into account
by means of an eigenvalue shift$^{24}$,
$\Delta V_{m} = - E^{3} L_{z} m_{l}$, for the 4{\it f} atom.
Here, L$_{z}$ is the z-component of the cerium total orbital moment,
m$_{l}$ is the magnetic quantum number, and 
E$^{3}$ is the Racah parameter evaluated 
self-consistently at each iteration.  

The crystalline field, which was neglected in 
the previous calculations,$^{4,5}$ is expected  
 to affect the magnetic behavior considerably,  
if it is large.
It is important to emphasize that
since in the first method the 4{\it f} states are treated as core
states,  they interact
 only with the spherical component of the effective
one-electron potential. Thus, the interaction of the {\it atomic-like}
4f state with the {\it non-spherical} components of the potential,
giving rise
to the CF splitting, $\Delta_{CF} = \epsilon_{\Gamma _8} -
\epsilon_{\Gamma _7}$, is not
 included in the calculation of the model Hamiltonian 
parameters. 
In this paper, we generalize the first, {\it ab initio} based, method
 to include both the interatomic 4{\it f}-4{\it f} coupling 
and the crystal-field interactions   
 on an equal footing and 
to employ a full potential LMTO evaluation
of the model Hamiltonian parameters.  While the effect of the full potential
on both the hybridization and coulomb exchange interactions
is small, including the CF interaction will
be shown to play a role as important as the 
interatomic 4{\it f}-4{\it f} interactions for
understanding and predicting the {\it overall trend}
 in the unusual magnetic properties,
as as one chemically tunes the degree of 4{\it f} localization across
this series of strongly correlated electron systems.
The resultant Hamiltonian is$^{4,5}$ 

\small   
\begin{eqnarray*}
{\it H}& =& -\sum_{i,j}\sum_{~^{\mu,\nu}_{\epsilon,\sigma}}
         \xi^{\epsilon \sigma}_{\mu \nu}(\theta_{ij})
         e^{-{\it i}(\mu-\nu+\epsilon-\sigma)\phi_{ij}}  
   c^{\dag}_{\epsilon}(j)c_{\sigma}(j)
         c^{\dag}_{\mu}(i)c_{\nu}(i) \\
   && + B_{4}\sum_{i}\left(O_{4}^{0}(i)+5O_{4}^{4}(i)\right) ,  
\end{eqnarray*}
\vspace{-30pt}
\begin{equation}
\end{equation}
\normalsize
where the ${\xi}^{\epsilon \sigma}_{\mu \nu}(\theta_{ij})$ are the
two-ion 4{\it f}-4{\it f} interaction matrices 
 rotated to a common crystal-lattice 
axis,
and the $O_{4}^{0}$ and
$O_{4}^{4}$ are the Stevens operators equivalents acting on the  
Ce$^{3+}$ free-ion 4$f_{5/2}$ multiplet.$^{25}$
The CF splitting is $\Delta_{CF}$ = 360$B_4$;
a positive $B_4$
value gives the $\Gamma_7$ ground state, which is 
 experimentally observed.$^{19}$
While our work in progress is aimed at evaluating the CF splitting 
on a wholly {\it ab initio} basis, in the absence of
 an {\it ab initio} value of the CF interaction in this
class of strongly correlated cerium systems,
the $\Delta_{CF}$ is
set to the experimental values listed in Table 3.$^{10,19}$

In Table~\ref{taba},
we present the calculated values of the zero-temperature
cerium magnetic moment from the FPLMTO electronic structure calculations.
Listed in the table are values both with and without the orbital polarization
correction taken into account. Note, the importance of including the
orbital polarization in these 4f correlated electron systems. 
As expected, in all
cases, the orbital polarization is found to be opposite to the spin
polarization. Comparison of the total energies predicts that
the magnetic anisotropy changes from the $<001>$ direction in the pnictides
to the $<111>$ in the chalcogenides, in agreement with experiment.
On the other hand, except perhaps for the lighter chalcogenides (CeS and CeSe),
comparison of the {\it ab initio} and experimental values for the
magnetic moment indicates the {\it failure} of the LDA calculations
to treat properly the correlation effects of the 4{\it f} states
(treated as valence states) within the LDA as the degree of 4{\it f} 
correlations increases in the heavier pnictide systems.
Furthermore, the {\it ab inito} calculations fail to predict
the large {\it moment collapse} from CeSb to CeTe, the latter being
described as an incipient heavy Fermion system.$^{2,3,20,21}$

\begin{table}[tbp]
\caption{
Values of the calculated and experimental $^{7-11}$ 
low-temperature ordered magnetic
moments for the cerium chalcogenides and pnictides in
units of $\mu_{B}$.  Listed are the LMTO values for
the spin moment $\mu_{S}$, the orbital
moment $\mu_{L}$, and total moment $\mu$, for the
spin polarized only calculation and for the calculation with
spin polarization and orbital polarization correction.
}
\begin{tabular}{|c|c|c|c|c|c|c|c|}
 & \multicolumn{2}{c}{ FP+SP } & &
    \multicolumn{2}{c}{ FP+SP+OP } & &
      EXPT \\
 & $\mu_{S}$ & $\mu_{L}$ & $\mu$ &  $\mu_{S}$ & $\mu_{L}$
 & $\mu$ &  $\mu$ \\ \hline
 CeS  & -1.00 & 0.91 &-0.09  &-1.24&1.99 & 0.75  &  0.57    \\
 CeSe & -1.08 & 1.02 &-0.06  &-1.26&2.07 & 0.81  &  0.57    \\
 CeTe & -1.15 & 1.28 & 0.07  &-1.31&2.29 & 0.98  &  0.30    \\ \hline
 CeP  & -0.80 & 0.55 &-0.25  &-0.85&1.27 & 0.43  &  0.80    \\
 CeAs & -0.84 & 0.64 &-0.20  &-0.85&1.42 & 0.57  &  0.80    \\
 CeSb & -0.86 & 0.74 &-0.12  &-0.91&1.61 & 0.70  &  2.06    \\
 CeBi & -0.86 & 0.74 &-0.12  &-0.95&1.69 & 0.74  &  2.10    \\ 
 \end{tabular}
 \label{taba}
 \end{table}

In Table~\ref{tabb}, we list the values of the $m$ = $m'$ =1/2 matrix elements
(characteristic matrix elements of the 6X6 exchange interaction matrix)
for the first three nearest-neighbor shells for the light (CeP and CeS)
and the heavier compounds (CeSb and CeTe). Listed separately in this table
are the three contributions to the interatomic 4{\it f}-4{\it f} 
interactions arising 
 from band-{\it f} hybridization
(V$^4$), band-{\it f} coulomb exchange (J$^2$), and the cross term.
It is important to note that while the coulomb exchange mediated interactions
dominate the magnetic behavior for the heavier, more localized,
4{\it f} systems, the opposite is true for the lighter, more delocalized,
 systems where the
hybridization mediated interactions dominate the magnetic behavior.
This change of behavior of the interatomic 
4{\it f}-4{\it f} interactions is a result
of the sensitivity of the hybridization and coulomb exchange to  
the degree of 4{\it f} localization. Equally important, is that
while both first and second nearest-neighbor 
  4{\it f}-4{\it f} interactions are ferromagnetic
for CeSb, there is an {\it interplay} between ferromagnetic 
first nearest-neighbor and antiferromagnetic second nearest-neighbor
interactions for CeTe.(These interactions are mediated via scattering
of conduction electrons). 
This results in a saturated ordered moment for CeSb and 
in the ordered {\it magnetic moment collapse}
 for CeTe (see Table III). 

In order to determine whether the magnetic moment collapse might be of
Kondo origin, we have evaluated the single-impurity Kondo temperature,$^{26}$ 
k$_B$T$_K$ = De$^{-\frac{1}{2\rho(E_F)|J(E_F)|}}$, 
across the series. Here, D is the bandwidth of the 
conduction electron states, $\rho(E_F)$ is the density of states of 
the conduction electrons at the Fermi energy, and $J(E_F)$ is the 
conduction electron-{\it f} exchange interaction at the Fermi energy, 
which has contributions   
 both from the  coulomb exchange interaction 
in Eq. (2), provided that it is negative, 
and the hybridization-induced exchange interaction  
 $|J_{hyb}(E_F)| =  \frac{V^2(E_F)U}{(|E_f - E_F|)(|E_f - E_F| + U)}$,  
 where $J_{hyb}(E_F) < 0$.
We find that the coulomb exchange interaction in Eq. (2) evaluated 
at the Fermi energy is positive across the entire series and hence
cannot give rise to the Kondo effect.
Thus, only the hybridization-induced exchange interaction, 
$J_{hyb}(E_F)$, 
can give rise to the Kondo effect.$^{26}$
Using the {\it ab initio} values of the parameters entering the 
expression for T$_K$, we find that T$_K \ll$ T$_{ord}$ across 
the entire series(T$_K <$ 10$^{-4}$K).
The Kondo temperatures in the 
monopnictide series is smaller than that in the chaclogenides, 
due to the fact that in the pnictides the Fermi 
energy lies in the pseudogap, resulting in low $\rho(E_F)$. 
These results suggest that  
 the moment collapse 
from CeSb to CeTe is of non-Kondo origin.
Rather, it results from an interplay of ferromagnetic and 
antiferromagnetic interatomic 4{\it f}-4{\it f} 
interactions which arises purely from differences 
in the underlying electronic structure.

\begin{table}[btp]
\caption{
Values of the $m$ = $m'$ =1/2 matrix elements 
(characteristic matrix elements of the 6X6 interatomic 
4{\it f}-4{\it f} interaction matrix
$E^{m_b m'_b}_{m_a m'_a}({\bf R_{b}-R_{a}})$, for the first, second, and 
third nearest-neighbor shells in degrees Kelvin. Listed are the values
of the hybridization
induced ($E_{V^{4}}$), cross terms ($E_{V^{2}J}$), and pure
coulomb exchange ($E_{J^{2}}$) contributions.}

\begin{tabular}{|c||r|r|r||r|r|r|}
  & \multicolumn{3}{c}{CeP} & \multicolumn{2}{r}{CeS}  &    \\  \hline
  & $E_{V^{4}}$ & $E_{V^{2}J}$ & $E_{J^{2}}$
  & $E_{V^{4}}$ & $E_{V^{2}J}$ & $E_{J^{2}}$                  \\ \hline
 R = ( $\frac{1}{2}$ $\frac{1}{2}$ 0 )
 &  2.23 &  0.64 &  1.53
 &  0.85 & -0.40 &  1.50
 \\ \hline
 R = ( 1 0 0 )
 &  6.39 &   0.27 &  1.65
 & -1.60 &   0.04 &  -0.80
 \\ \hline
 R = ( 1 $\frac{1}{2}$ $\frac{1}{2}$ )
 & -0.08 & -0.02 &  0.16
 &  0.38 & -0.16 &  0.13 \\ 
\end{tabular}

\begin{tabular}{|c||r|r|r||r|r|r|} 
  & \multicolumn{3}{c}{CeSb} & \multicolumn{2}{r}{CeTe}  &    \\  \hline
  & $E_{V^{4}}$ & $E_{V^{2}J}$ & $E_{J^{2}}$
  & $E_{V^{4}}$ & $E_{V^{2}J}$ & $E_{J^{2}}$                  \\ \hline
 R = ( $\frac{1}{2}$ $\frac{1}{2}$ 0 )
 &  0.70 &  0.34 &   7.30
 &  0.17 & -0.19 &   2.90
 \\ \hline
 R = ( 1 0 0 )
 &  2.07 &   0.07 &  10.21
 & -0.19 &   0.04 & -1.69
 \\ \hline
 R = ( 1 $\frac{1}{2}$ $\frac{1}{2}$ )
 & -0.02 & -0.03 &  0.40
 &  0.04 & -0.06 & -0.01
 \\
\end{tabular}
\label{tabb}
\end{table}

Listed in Table~\ref{tabc}  
are the calculated zero-temperature ordered moment and
ordering temperature, T$_N$, from the first, {\it ab initio} based, method,
with and without the CF interaction. It is clear that for the heavier
systems (CeBi, CeSb, CeTe) the effect of the CF interaction on the
magnetic moments is small, and it is slightly more pronounced on the
ordering temperatures. This is due to the fact that for the more localized
systems the CF interaction is smaller than the two-ion interactions.
This is the reason that the previous calculations,$^5$ neglecting the CF
interaction, gave results in very good agreement with experiment.
On the other hand, for the lighter more delocalized systems the CF
interactions are much larger than the interatomic 
4{\it f}-4{\it f} interactions, 
and hence dominate the magnetic behavior. The overall decrease of the magnetic
moments in the presence of the CF interaction in all systems,
arises from the mixing
of the off-diagonal angular momentum states $|\pm5/2>$ and $|\mp3/2>$ states
from the CF interaction  with $\Gamma_7$ ground state.
Overall, we find that 
the first, {\it ab initio} based, approach 
which takes into account  all three pertinent interactions 
(hybridization, coulomb exchange, and CF
interactions) on an equal footing, 
yields results 
for both the zero-temperature moment and the
ordered temperature (a more stringent test for the theory)
in excellent agreement with experiment.

A final corroboration of the success of the first {\it ab initio} based
method is that it predicts the experimentally observed 
change of the ground-state magnetic 
structure from the  
$<001>$ antiferromagnetic
type IA ($\uparrow \uparrow \downarrow \downarrow)$ in CeBi and CeSb
 to the $<001>$ antiferromagnetic type I ($\uparrow \downarrow$) 
 in CeAs and CeP.$^{7,15}$ More specifically, the sign [ferromagnetic (F) or 
antiferromagnetic (AF)] of the 
$|\pm 5/2>$ matrix elements of the 6X6 exchange matrix determines 
the {\it interplanar} interaction between successive (001) Ce planes.  
We find, that for the heavier compounds (CeBi and CeSb) 
the $|\pm 5/2>$ matrix elements of the {\it coulomb exchange} matrix 
are FM  and hence favor the $\uparrow \uparrow \downarrow \downarrow$ 
type, while in the lighter systems (CeAs and CeP) the 
$|\pm 5/2>$ matrix elements of the {\it hybridization}-induced two-ion matrix 
are AF, and hence they favor the $\uparrow \downarrow$ type.

\begin{table}[tbp]
\caption{
Calculated  values (from the first {\it ab initio} based method)
of the zero-temperature ordered moment $\mu$ in $\mu_{B}$, 
and the ordering temperature
$T_{N}$ in degrees Kelvin, 
with and without the crystalline field (CF) interaction 
across the cerium pnictide and chalcogenides series. Also listed 
are the experimental values$^{7-11}$ of $\mu$ and $T_{N}$, and the CF 
splitting$^{19}$ between the $\Gamma_7$ ground state and the $\Gamma_8$ state 
in degrees Kelvin.}
\begin{tabular}{|d|d|c|c|c|c|c|c|d|}  
 & $\Delta_{CF}$ &
   \multicolumn{2}{r}{ $\mu_{0}$ }& & &
   \multicolumn{2}{r}{ $T_{N}$ }     &    \\ \hline
 & & no CF & CF & exp  &  & no CF & CF & exp \\ \hline
CeS  & 140 & 1.80 & 0.73 & 0.57 & & 1.0 & 11.0 & 8.4 \\
CeSe & 116 & 1.10 & 0.79 & 0.57 & & 2.5 & 14.0 & 5.7 \\
CeTe &  32 & 0.60 & 0.46 & 0.30 & & 8.0 &  5.0 & 2.2 \\ \hline
CeP  & 150 & 2.10 & 0.73 & 0.81 & & 14 & 11 & 8 \\
CeAs & 137 & 2.10 & 0.74 & 0.85 & & 16 & 13 & 8 \\
CeSb &  37 & 2.10 & 1.80 & 2.06 & & 20 & 18 & 17 \\
CeBi &   8 & 2.10 & 2.10 & 2.10 & & 40 & 40 & 26 \\  
\end{tabular}
\label{tabc}
\end{table}

In conclusion, we have applied two different, {\it ab inito} based
and {\it ab initio} LDA, 
methods to study the dramatic change of magnetic properties across
a series of strongly correlated electron systems which offer the
opportunity to chemically tailor the different type of interactions
(band-{\it f} hybridization, band-{\it f} coulomb exchange, 
and CF interactions),
pertinent to the unusual magnetic behavior. 
The first, {\it ab initio} based, approach which explicitly takes into account
the interplay of the three pertinent interactions, gives results in excellent
agreement with experiment for all compounds in the series, including
the {\it moment collapse} from CeSb to CeTe and the trend of moments
and ordering temperatures across the series.
The remaining problem of determining on a wholly {\it ab initio}
basis the {\it suppressed} crystal-field interactions in this
class of systems poses a theoretical challenge for future theoretical
work.
On the other hand, the second, fully {\it ab initio} LDA, method
gives good results for the lighter chalcogenide systems, but it entirely
fails to give, even qualitatively, the trend of the unusual magnetic behavior.

The research at California State University Northridge (CSUN) was
supported by the National Science Foundation under Grant No.
DMR-9531005, by the US Army Grant No. DAAH04-95, and the CSUN Office
of Research and Sponsored Projects,
and at West Virginia University by the NSF under Grant No.
DMR-9120333.

\end{document}